\documentclass[prd,preprint,nofootinbib,a4paper,11pt]{revtex4}

\usepackage[centertags]{amsmath}
\usepackage{amsfonts}
\usepackage{amssymb}
\usepackage[sort&compress]{natbib}
\usepackage[hypertex]{hyperref}

\DeclareMathOperator{\Sp}{Sp}

\begin{document}

\title{Comment on ``Dual path integral representation for finite temperature quantum field theory''}

\date{\today}

\author{P.O. Kazinski}

\email{kpo@phys.tsu.ru}

\affiliation{Physics Faculty, Tomsk State University, Tomsk, 634050 Russia}

\begin{abstract}

I show that the novel dual path integral representation for finite temperature quantum field theory proposed in [Phys. Rev. D \textbf{77}, 105030 (2008)] is a well-known representation of quantum mechanics in terms of symbols of operators.

\end{abstract}

\maketitle

The formulation of quantum mechanics in terms of functions on the phase-space of an underlying classical system was already considered by Moyal \cite{Moyal}. The striking feature of this representation is that the classical limit of quantum dynamics becomes evident. In that limit, the quantum Liouville (von Neumann) equation passes into the classical Liouville equation, symbols of quantum operators turn out to be classical observables on the phase-space, the Wigner function, that is a symbol of the density operator, becomes the Liouville probability density function. Besides, in such a representation, a trace of an operator is simply an integral of the corresponding symbol over the phase-space. The paper \cite{CTFMR} illustrates some of these properties of the symbol representation of quantum mechanics once again in the path integral framework. Notice that a path integral representation of symbols of operators is also well-known (see, e.g., \cite{BerezMSQ}).

Now we shall see that the partition function in the dual description \cite{CTFMR} is a  representation of the trace of the (unnormalized) density operator by means of its symbol. Let us break the interval of integration in Eq. (16) of \cite{CTFMR} over $\mathcal{DQ}$ like in formula (7), i.e., from $-T$ to $0$, from $0$ to $\beta$ and from $\beta$ to $T$. In the limit $T\rightarrow+\infty$, similarly to Eq. (10) of \cite{CTFMR} we have
\begin{equation}
    \mathcal{Z}_s(\beta)=e^{\beta E_0}\int\frac{d\xi_1d\xi_2}{(2\pi)^2}\langle0|\int \mathcal{DQ}e^{-\mathcal{S}(\mathcal{Q})+i\int_0^\beta d\tau j_a(\tau)\mathcal{Q}_a(\tau)}|0\rangle,
\end{equation}
where the integral in the action functional $\mathcal{S}(\mathcal{Q})$ is taken over the time interval $[0,\beta]$. Returning to operators we obtain
\begin{multline}\label{part func}
    \mathcal{Z}_s(\beta)=\int\frac{d\xi_1d\xi_2}{2\pi}\langle0|e^{-\beta(H(\hat{p}+\xi_1,\hat{q}+\xi_2)-E_0)}|0\rangle\\
    =\int\frac{d\xi_1d\xi_2}{2\pi}\langle0|e^{-i\xi_1\hat{q}}e^{i\xi_2\hat{p}}e^{-\beta(H(\hat{p},\hat{q})-E_0)}e^{-i\xi_2\hat{p}}e^{i\xi_1\hat{q}}|0\rangle=\Sp e^{-\beta(H(\hat{p},\hat{q})-E_0)}.
\end{multline}
The integrand is exactly the symbol of the density operator. Such symbols were thoroughly studied in \cite{Klauder,McKenKlau}. Furthermore, for a free model, when the Hamiltonian $\hat{H}$ is quadratic, the integrand function is nothing but the Wick symbol of the density operator \cite{BerezMSQ,Klauder,McKenKlau,Bargmann}. The periodicity condition for both phase-space variables in the path integral representation of the trace of the density operator simply follows from \eqref{part func} or the trace formula (6.12) of \cite{McKenKlau} and is pointed out, for instance, in Eq. (5.6) of \cite{BerezMSQ}.

\end{document}